\newif\ifANON{}
\newif\ifACM{}
\newif\ifLNCS{}
\newif\ifIACRTRANS{}
\newif\ifIEEE{}
\newif\ifUSENIX{}
\newif\ifSREP{}
\newif\ifSOFTWAREX{}
\newif\ifLINUXBUILD{}

\LNCStrue{}

\IfFileExists{.tmp.anon.tmp}{\ANONtrue}{}
\IfFileExists{/dev/null}{\LINUXBUILDtrue}{}

\ifACM{}
\ifANON{}
\documentclass[sigconf,review=true,anonymous=true,nonacm=true]{acmart}
\else
\documentclass[sigconf,review=true,anonymous=true,nonacm=true]{acmart}
\fi
\author{Arttu Paju}
\orcid{0000-0002-1398-2184}
\affiliation{
\institution{Tampere University}
\city{Tampere}
\country{Finland}
}
\email{arttu.paju@tuni.fi}

\author{Juha Nurmi}
\orcid{0000-0003-3071-9027}
\affiliation{
\institution{Tampere University}
\city{Tampere}
\country{Finland}
}
\email{juha.nurmi@tuni.fi}

\author{Alejandro Cabrera Aldaya}
\orcid{0000-0002-1544-6772}
\affiliation{
\institution{Tampere University}
\city{Tampere}
\country{Finland}
}
\email{alejandro.cabreraaldaya@tuni.fi}

\author{Nicola Tuveri}
\orcid{0000-0001-5172-4568}
\affiliation{
\institution{Tampere University}
\city{Tampere}
\country{Finland}
}
\email{nicola.tuveri@tuni.fi}

\author{Juha Savim\"{a}ki}
\orcid{0000-0002-5252-7774}
\affiliation{
\institution{Tampere University}
\city{Tampere}
\country{Finland}
}
\email{juha.savimaki@tuni.fi}

\author{Marko Kivikangas}
\orcid{0009-0002-6990-0202}
\affiliation{
\institution{Tampere University}
\city{Tampere}
\country{Finland}
}
\email{marko.kivikangas@tuni.fi}

\fi

\ifLNCS{}
\documentclass[runningheads]{llncs}
\usepackage{xcolor}
\usepackage[pdftex,colorlinks=true,pdfstartview=FitV,linkcolor=black,citecolor=blue,urlcolor=black]{hyperref}

\usepackage[square,comma,numbers,sectionbib]{natbib}

\makeatletter
\renewcommand\@biblabel[1]{#1.}
\makeatother
\ifANON\else
\renewcommand\orcidID[1]{}

\author{
Arttu~Paju\inst{1}\orcidID{0000-0002-1398-2184} \and
Juha~Nurmi\inst{1}\orcidID{0000-0003-3071-9027} \and
Alejandro~Cabrera~Aldaya\inst{1}\orcidID{0000-0002-1544-6772}
\and Nicola~Tuveri\inst{1}\orcidID{0000-0001-5172-4568} \and
Juha~Savim\"{a}ki\inst{1,2}\orcidID{0000-0002-5252-7774} \and
Marko Kivikangas\inst{1}\orcidID{0009-0002-6990-0202} \and
Brian~McGillion\inst{3}\orcidID{0000-0003-2683-5295}
}

\institute{
 Tampere University, Tampere, Finland,
 \email{\{arttu.paju,juha.nurmi,alejandro.cabreraaldaya,nicola.tuveri,juha.savimaki\}@tuni.fi}
  \and
  Unikie Oy, Tampere, Finland
  \and
  Technology Innovation Institute (TII), Abu Dhabi, UAE, \email{brian.mcgillion@tii.ae}
}
\fi
\fi

\ifIACRTRANS{}
\ifANON{}
\documentclass[journal=tches,submission]{iacrtrans}
\else
\documentclass[journal=tches,preprint]{iacrtrans}
\fi
\usepackage[backend=bibtex,style=alphabetic,maxbibnames=99,natbib=true]{biblatex}
\author{Arttu~Paju\inst{1} \and Juha~Nurmi\inst{1} \and Alejandro~Cabrera~Aldaya\inst{1}
\and Nicola~Tuveri\inst{1} \and Juha~Savim\"{a}ki\inst{1,2} \and Brian~McGillion\inst{3}
\and Marko Kivikangas\inst{1}}
\institute{
 Tampere University, Tampere, Finland,
 \email[arttu.paju@tuni.fi,juha.nurmi@tuni.fi,alejandro.cabreraaldaya@tuni.fi,nicola.tuveri@tuni.fi,juha.savimaki@tuni.fi,brian@ssrc.tii.ae,marko.kivikangas@tuni.fi]{
 {arttu.paju,juha.nurmi,alejandro.cabreraaldaya,nicola.tuveri,juha.savimaki,marko.kivikangas}@tuni.fi}
  \and
  Unikie Oy, Tampere, Finland
  \and
  Technology Innovation Institute (TII), Abu Dhabi, UAE, \email{brian.mcgillion@tii.ae}
}

\fi

\ifIEEE{}
\documentclass[10pt,conference,compsocconf]{IEEEtran}
\usepackage[square,sort&compress,numbers]{natbib}

\ifANON\else
\author{\IEEEauthorblockN{Alejandro Cabrera Aldaya\IEEEauthorrefmark{1},
Billy Bob Brumley\IEEEauthorrefmark{2},
Sohaib ul Hassan\IEEEauthorrefmark{2},
Cesar Pereida Garc\'{\i}a\IEEEauthorrefmark{2},
Nicola Tuveri\IEEEauthorrefmark{2}}
\IEEEauthorblockA{\IEEEauthorrefmark{1}Universidad Tecnol\'ogica de la Habana (CUJAE), Habana, Cuba}
\IEEEauthorblockA{\IEEEauthorrefmark{2}Tampere University, Tampere, Finland}}
\fi
\fi

\ifUSENIX{}
\documentclass[letterpaper,twocolumn,10pt]{article}
\makeatletter
\@namedef{ver@breakurl.sty}{}%
\@namedef{ver@cite.sty}{}%
\makeatother
\usepackage[hyphens]{url} %
\usepackage{usenix}
\usepackage[sort,numbers]{natbib}
\date{}
\ifANON\else
\author{
\textnormal{Your N.\ Here}\\
Your Institution
\and
\textnormal{Second Name}\\
Second Institution
} %

\newcommand{\authormark}[1]{\textsuperscript{#1}}
\author{
\textnormal{Cesar~Pereida~Garc{\'{\i}}a\authormark{1}, Sohaib~ul~Hassan\authormark{1}, Nicola~Tuveri\authormark{1},}\\
\textnormal{Iaroslav~Gridin\authormark{1}, Alejandro~Cabrera~Aldaya\authormark{1,2}, and Billy~Bob~Brumley\authormark{1}}\\
\authormark{1}Tampere University, Tampere, Finland\\
\{cesar.pereidagarcia,n.sohaibulhassan,nicola.tuveri,iaroslav.gridin,billy.brumley\}@tuni.fi\\
\authormark{2}Universidad Tecnol\'ogica de la Habana (CUJAE), Habana, Cuba\\
aldaya@gmail.com
}

\fi
\fi

\ifSREP{}
\documentclass[fleqn,10pt]{wlscirep}
\usepackage[utf8]{inputenc}

\providecommand\citep{}
\renewcommand{\citep}[1]{\cite{#1}}
\providecommand\citet{}
\renewcommand{\citet}[1]{\cite{#1}}

\ifANON{}
\author[]{Anonymous authors}
\else
\author[1,*]{Juha Nurmi}
\author[1]{Arttu Paju}
\author[2]{Billy Bob Brumley}

\affil[1]{Tampere University, Tampere, FI-33720, Finland}
\affil[2]{Rochester Institute of Technology, Rochester, 14623-5608, USA} %

\affil[*]{juha.nurmi@tuni.fi}
\fi
\fi

\ifSOFTWAREX{}
\documentclass[preprint,12pt,a4paper]{elsarticle}
\usepackage[utf8]{inputenc}
\usepackage{amssymb}
\usepackage{lineno}
\journal{SoftwareX}
\fi

\usepackage[T1]{fontenc}
\usepackage{graphicx}
\usepackage{xspace}
\usepackage{lipsum}
\usepackage{multirow}
\usepackage{array} %
\usepackage{booktabs} %
\usepackage{hyperref} %
\usepackage{url} %
\usepackage{array} %
\usepackage{placeins} %
\usepackage{subcaption} %

\usepackage{tikz}

\usepackage[shortcuts]{glossaries}
\glsdisablehyper{}

\usepackage{xcolor}
\usepackage{tcolorbox}

\hypersetup{
    colorlinks=true,
    linkcolor=blue,
    filecolor=magenta,
    urlcolor=cyan,
    citecolor=blue
}

\newcommand{\TITLE}{External entropy supply for IoT devices employing a RISC-V Trusted Execution Environment}

\newcommand{\CPU}{\gls{CPU}}
\newcommand{\SDK}{\gls{SDK}}
\newcommand{\SPIRS}{\gls{SPIRS}}
\newcommand{\RISCV}{RISC-V}
\newcommand{\RoT}{\gls{RoT}}
\newcommand{\TEE}{\gls{TEE}}
\newcommand{\IoT}{\gls{IOT}}
\newcommand{\PUF}{\gls{PUF}}
\newcommand{\SM}{\gls{SM}}
\newcommand{\TA}{\gls{TA}}
\newcommand{\TAs}{\glspl{TA}}
\newcommand{\CA}{\gls{CA}}
\newcommand{\CAs}{\glspl{CA}}
\newcommand{\GP}{\gls{GP}}
\newcommand{\PMP}{\gls{PMP}}
\newcommand{\EaaS}{\gls{EaaS}}
\newcommand{\TRNG}{\gls{TRNG}}
\newcommand{\TRNGs}{\glspl{TRNG}}

\newcommand{\RNGs}{\glspl{RNG}}
\newcommand{\CSPRNG}{\gls{CSPRNG}}
\newcommand{\PRNG}{\gls{PRNG}}
\newcommand{\PRNGs}{\glspl{PRNG}}
\newcommand{\TES}{\gls{TES}}
\newcommand{\HTTP}{\gls{HTTP}}
\newcommand{\TTP}{\gls{TTP}}
\newcommand{\NTP}{\gls{NTP}}
\newcommand{\TCB}{\gls{TCB}}

\newcommand{\DoS}{\gls{DoS}}
\newcommand{\PKI}{\gls{PKI}}
\newsavebox{\discard}

\usepackage{lettrine}

\ifLNCS{}
\newcommand{\Paragraph}[2][.]{\medbreak\noindent\textbf{#2#1}}
\fi
\ifIACRTRANS{}
\newcommand{\Paragraph}[2][.]{\paragraph{#2#1}}
\fi
\ifUSENIX{}
\newcommand{\Paragraph}[2][.]{\smallbreak\noindent\textbf{#2#1}}
\fi
\ifACM{}
\newcommand{\Paragraph}[2][.]{\subsubsection*{#2#1}}
\fi
\ifIEEE{}
\makeatletter
\renewcommand{\@IEEEsectpunct}{~}
\makeatother
\newcommand{\Paragraph}[2][.]{\subsubsection*{#2#1}}
\fi

\newacronym{CPU}{CPU}{central processing unit}
\newacronym{RAM}{RAM}{random access memory}
\newacronym{GPU}{GPU}{graphics processing unit}

\newacronym{SDK}{SDK}{software development kit}
\newacronym{SPIRS}{SPIRS}{Secure Platform for ICT Systems Rooted at the Silicon Manufacturing Process}
\newacronym{IMA}{IMA}{Integrity Measurement Architecture}
\newacronym{IOT}{IoT}{Internet of Things} 
\newacronym{ICT}{ICT}{Information and Communications Technology}
\newacronym{RoT}{RoT}{Root of Trust}
\newacronym{TEE}{TEE}{Trusted Execution Environment}
\newacronym{HLOS}{HLOS}{High-Level Operating System}
\newacronym{PUF}{PUF}{Physical Unclonable Function}
\newacronym{SM}{SM}{Security Monitor}
\newacronym{TA}{TA}{Trusted Application}
\newacronym{CA}{CA}{Client Application}
\newacronym{GP}{GP}{GlobalPlatform}
\newacronym{PMP}{PMP}{Physical Memory Protection}
\newacronym{EaaS}{EaaS}{Entropy as a Service}
\newacronym{TRNG}{TRNG}{True Random Number Generator}
\newacronym{RNG}{RNG}{Random Number Generator}
\newacronym{PRNG}{PRNG}{Pseudorandom Number Generator}
\newacronym{CSPRNG}{CSPRNG}{Cryptographically Secure Pseudorandom Number Generator}
\newacronym{TES}{TES}{Trusted Execution Server}
\newacronym{HTTP}{HTTP}{Hypertext Transfer Protocol}
\newacronym{TTP}{TTP}{Trusted Third Party}
\newacronym{TLS}{TLS}{Transport Layer Security}
\newacronym{NTP}{NTP}{Network Time Protocol}
\newacronym{TCB}{TCB}{Trusted Computing Base}
\newacronym{OS}{OS}{Open Source}
\newacronym{DoS}{DoS}{Denial of Service}

\newacronym{PKI}{PKI}{Public-Key Infrastructure}
\newacronym{PQC}{PQC}{Post-Quantum Cryptography}
\newacronym{PQ/T}{PQ/T}{Post-Quantum/Traditional}
\newacronym{HNDL}{HNDL}{Harvest-Now-Decrypt-Later}
\newacronym{CRQC}{CRQC}{Cryptographically Relevant Quantum Computer}

\newcommand{\KEYWORDS}{%
RISC-V hardware \and
Entropy \and
Randomness \and
Trusted Execution Environment \and
Cryptography \and
Network protocols \and
IoT %
}

\title{\TITLE{}} %

\ifACM{}
\begin{abstract}
Entropy---a measure of randomness---is compulsory for the generation of secure cryptographic keys;
however, \IoT{} devices %
that are small or constrained often struggle to collect sufficient entropy.
In this article, we solve the entropy provisioning problem for a fleet of \IoT{} devices that can generate a limited amount of entropy.
We employ a \TEE{} based on \RISCV{} to create an external entropy service for a fleet of \IoT{} devices.
A small measure of true entropy or pre-installed keys can establish initial secure communication.
Once connected, devices can request cryptographically strong entropy from a \TEE{}-backed server.
\RISCV{} offers \TRNGs{} and a \TEE{} for devices to attest that they are receiving reliable entropy.
In addition, this solution can be expanded by adding \IoT{} devices with sensors
that produce high-quality entropy as additional entropy sources for the \RISCV{} entropy provider.
Our open-source implementation shows that building trusted entropy infrastructure for \IoT{} is both feasible and effective on open \RISCV{} platforms.
 \end{abstract}
\fi

\ifSREP{}
\begin{abstract}
Entropy---a measure of randomness---is compulsory for the generation of secure cryptographic keys;
however, \IoT{} devices %
that are small or constrained often struggle to collect sufficient entropy.
In this article, we solve the entropy provisioning problem for a fleet of \IoT{} devices that can generate a limited amount of entropy.
We employ a \TEE{} based on \RISCV{} to create an external entropy service for a fleet of \IoT{} devices.
A small measure of true entropy or pre-installed keys can establish initial secure communication.
Once connected, devices can request cryptographically strong entropy from a \TEE{}-backed server.
\RISCV{} offers \TRNGs{} and a \TEE{} for devices to attest that they are receiving reliable entropy.
In addition, this solution can be expanded by adding \IoT{} devices with sensors
that produce high-quality entropy as additional entropy sources for the \RISCV{} entropy provider.
Our open-source implementation shows that building trusted entropy infrastructure for \IoT{} is both feasible and effective on open \RISCV{} platforms.
 \end{abstract}
\fi

\begin{document}

\ifSREP{}
\flushbottom
\fi

\ifSOFTWAREX
\else
\maketitle
\fi

\ifLNCS{}
\begin{abstract}
Entropy---a measure of randomness---is compulsory for the generation of secure cryptographic keys;
however, \IoT{} devices %
that are small or constrained often struggle to collect sufficient entropy.
In this article, we solve the entropy provisioning problem for a fleet of \IoT{} devices that can generate a limited amount of entropy.
We employ a \TEE{} based on \RISCV{} to create an external entropy service for a fleet of \IoT{} devices.
A small measure of true entropy or pre-installed keys can establish initial secure communication.
Once connected, devices can request cryptographically strong entropy from a \TEE{}-backed server.
\RISCV{} offers \TRNGs{} and a \TEE{} for devices to attest that they are receiving reliable entropy.
In addition, this solution can be expanded by adding \IoT{} devices with sensors
that produce high-quality entropy as additional entropy sources for the \RISCV{} entropy provider.
Our open-source implementation shows that building trusted entropy infrastructure for \IoT{} is both feasible and effective on open \RISCV{} platforms.
\end{abstract}
\fi

\ifIACRTRANS{}
\begin{abstract}
Entropy---a measure of randomness---is compulsory for the generation of secure cryptographic keys;
however, \IoT{} devices %
that are small or constrained often struggle to collect sufficient entropy.
In this article, we solve the entropy provisioning problem for a fleet of \IoT{} devices that can generate a limited amount of entropy.
We employ a \TEE{} based on \RISCV{} to create an external entropy service for a fleet of \IoT{} devices.
A small measure of true entropy or pre-installed keys can establish initial secure communication.
Once connected, devices can request cryptographically strong entropy from a \TEE{}-backed server.
\RISCV{} offers \TRNGs{} and a \TEE{} for devices to attest that they are receiving reliable entropy.
In addition, this solution can be expanded by adding \IoT{} devices with sensors
that produce high-quality entropy as additional entropy sources for the \RISCV{} entropy provider.
Our open-source implementation shows that building trusted entropy infrastructure for \IoT{} is both feasible and effective on open \RISCV{} platforms.
\end{abstract}
\fi

\ifIEEE{}
\begin{abstract}
Entropy---a measure of randomness---is compulsory for the generation of secure cryptographic keys;
however, \IoT{} devices %
that are small or constrained often struggle to collect sufficient entropy.
In this article, we solve the entropy provisioning problem for a fleet of \IoT{} devices that can generate a limited amount of entropy.
We employ a \TEE{} based on \RISCV{} to create an external entropy service for a fleet of \IoT{} devices.
A small measure of true entropy or pre-installed keys can establish initial secure communication.
Once connected, devices can request cryptographically strong entropy from a \TEE{}-backed server.
\RISCV{} offers \TRNGs{} and a \TEE{} for devices to attest that they are receiving reliable entropy.
In addition, this solution can be expanded by adding \IoT{} devices with sensors
that produce high-quality entropy as additional entropy sources for the \RISCV{} entropy provider.
Our open-source implementation shows that building trusted entropy infrastructure for \IoT{} is both feasible and effective on open \RISCV{} platforms.
 \end{abstract}
\fi

\ifUSENIX{}
\begin{abstract}
Entropy---a measure of randomness---is compulsory for the generation of secure cryptographic keys;
however, \IoT{} devices %
that are small or constrained often struggle to collect sufficient entropy.
In this article, we solve the entropy provisioning problem for a fleet of \IoT{} devices that can generate a limited amount of entropy.
We employ a \TEE{} based on \RISCV{} to create an external entropy service for a fleet of \IoT{} devices.
A small measure of true entropy or pre-installed keys can establish initial secure communication.
Once connected, devices can request cryptographically strong entropy from a \TEE{}-backed server.
\RISCV{} offers \TRNGs{} and a \TEE{} for devices to attest that they are receiving reliable entropy.
In addition, this solution can be expanded by adding \IoT{} devices with sensors
that produce high-quality entropy as additional entropy sources for the \RISCV{} entropy provider.
Our open-source implementation shows that building trusted entropy infrastructure for \IoT{} is both feasible and effective on open \RISCV{} platforms.
 \end{abstract}
\fi

\glsresetall{}

\section{Introduction}\label{sec:intro}

Entropy---a measure of randomness---is compulsory for the generation of secure cryptographic keys.
Small or resource-constrained \IoT{} devices %
often struggle to collect sufficient high-quality entropy for secure cryptographic operations.
When constrained devices experience entropy starvation, it can result in predictable cryptographic outputs that undermine the security of keys and communications.

Previously, this problem was solved with
(i) \TRNGs{}, which use inherently random physical processes like electronic noise
or photonic processes to provide a built-in entropy source, which is usually part of the device's main processor;
(ii) Hybrid Entropy Sources, which use both software (pseudo-random) and hardware (true random) sources to produce random values; and
(iii) Environmental Sources, which use \IoT{} sensors, such as temperature, light, and motion sensors, can serve as sources of entropy if their readings are unpredictable;
instability from oscillators or clock drift can also be a source of entropy.
User interaction, i.e., a user touching buttons or moving the device as a source of entropy, is an extension of this concept.

However, it is important to take precautions to prevent adversaries from easily manipulating environmental data.
If these options are unavailable or provide a limited amount of entropy,
techniques such as whitening algorithms~\citep{DBLP:journals/corr/abs-2208-11935}
can be used to derive longer keys from lower-entropy sources.
Furthermore---using extractors such as cryptographic hash functions or key derivation functions---a small quantity of entropy
that is available as a seed can be used to generate a larger, deterministic, but pseudo-random
value~\citep{VonNeumann, DBLP:conf/focs/SanthaV84, DBLP:conf/crypto/Krawczyk10}.
However, these techniques may be computationally intensive for devices with limited resources.

Another option is to provide unique device certificates during manufacturing or configuration,
but even if an \IoT{} device begins with a strong key,
that key must be periodically regenerated or re-seeded with new entropy to maintain security.

In this article, we solve the problem of entropy provisioning for a fleet of \IoT{} devices that can generate a limited amount of entropy:
A \TEE{} based on \RISCV{} provides entropy for a fleet of devices.
Initial secure communication can be established with a small measure of true entropy or with pre-installed keys.
After establishing the channel, devices receive entropy from a reliable external source.
\RISCV{}-based \SPIRS{} project provides \TRNGs{} and \TEE{} mechanisms for devices to attest to the provision of entropy.
In addition, this solution can be expanded by adding \IoT{} devices with sensors that produce
entropy as additional entropy sources for the \RISCV{} entropy provider.
\TEE{} is used to enable a design with no \TTP{} in our \EaaS{} protocol.

We make the following novel contributions in this paper:
(i) We design, develop, and publish a \SPIRS{} \TEE{} \SDK{} that enables the development of \TAs{}
for the \RISCV{}-based \SPIRS{} platform---allowing open-source development of \TAs{} for a \RISCV{}-based \TEE{}.
(ii) We design, develop, and publish an \EaaS{} \TA{} using the \SPIRS{} \TEE{} \SDK{}---demonstrating
how a \RISCV{}-based \TEE{} can be used to create a secure external entropy service for a fleet of \IoT{} devices.
\section{Background}\label{sec:background}

Entropy generates the desired randomness for security services.
In practice, this randomness must be generated artificially.
Achieving true randomness can be challenging and requires careful design.
The aim of this research is to identify a solution to provide true entropy for a fleet of \IoT{} devices that lack the resources to generate it themselves.

\RISCV{} is a reduced instruction set computer (RISC) architecture that is used to develop custom processors for many different purposes.
\TEE{} hardware is a high-level security solution to create a secure area for the \CPU{}.
This isolated security area aims to assure confidentiality and integrity.
To achieve the aim of this research, \RISCV{} and \TEE{} are both used within \SPIRS{}.
The \SPIRS{} project is introduced more thoroughly in
\autoref{subsec:spirs_hw_platform}, \autoref{subsec:spirs_tee_platform}, and \autoref{subsec:spirs_sw}.

\subsection{Hardware: \SPIRS{} Platform}\label{subsec:spirs_hw_platform}

\begin{figure}[tb]
  \begin{center}
  \includegraphics[width=1.0\textwidth]{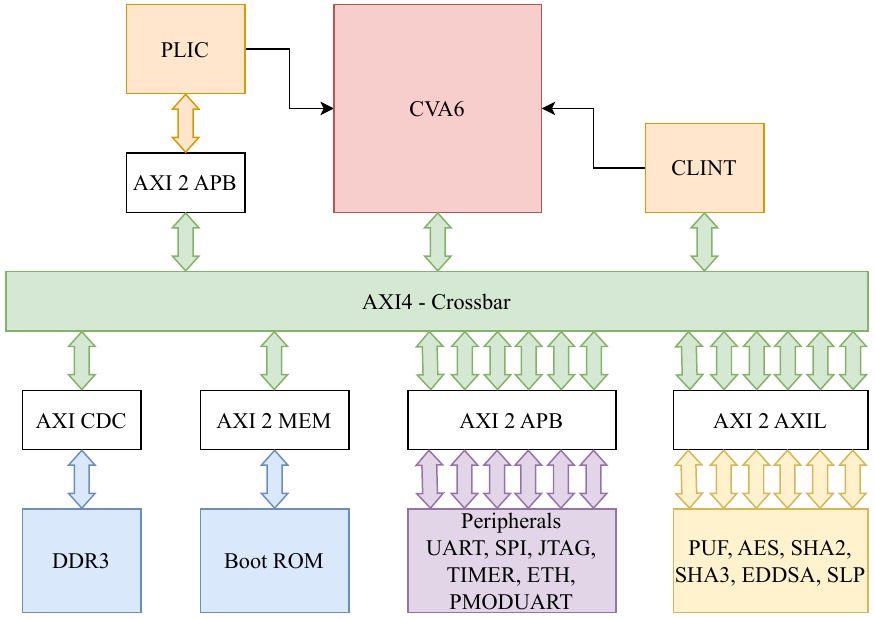}
  \end{center}
  \caption{\centering{\SPIRS{} Hardware Platform}}%
  \label{fig_spirs_platform}
\end{figure}

The \SPIRS{}~\citep{spirs_web} EU-funded project addresses innovative
approaches to provide security and data privacy to \gls{ICT} elements.
The project encompasses the complete design of a \SPIRS{}
platform, which integrates a dedicated hardware \RoT{}, a \RISCV{}
processor core, and some peripherals.
The \SPIRS{} hardware platform
(\autoref{fig_spirs_platform})
includes three components:

\begin{description}
    \item[CORE-V CVA6 Core:] The \RISCV{} processor in \SPIRS{} is
          based on the open-source CORE-V CVA6 core, developed by the
          OpenHW group.
          The \SPIRS{} project enhances the security of this core by
          addressing vulnerabilities identified in the
          PP84 protection profile of the Common Criteria~\citep{CCPP84}.
          Security countermeasures have been applied to
          ensure the processor operates securely, even in untrusted
          environments.

    \item[\RoT{}:] The \RoT{} is the foundational
          security element of the \SPIRS{} platform, as it is the source
          of trust for the entire system built over it.
          The security of software components (execution environment, boot
          process, applications) relies on identifiers, random numbers,
          and cryptographic functions that are provided by the \RoT{}.
          The \RoT{} is modular and flexible, allowing for the addition or
          removal of components to optimize the system for specific
          applications.
          This flexibility ensures that the platform can adapt to new
          security threats by updating its \RoT{}.
          The current version of the hardware \RoT{} includes:
          \begin{description}
              \item[\PUF{}/\TRNG{}:] A hardware-based
                    mechanism, based on a Ring Oscillator architecture,
                    which generates unique identifiers and random
                    numbers for device authentication and security
                    operations.
              \item[AES-256:] A symmetric encryption engine
                    for securing data.
              \item[SHA-256:] A cryptographic hashing function used
                    to ensure data integrity.
              \item[Digital Signatures:] A hardware implementation
                    of algorithms to accelerate and secure digital
                    authentication.
          \end{description}

    \item[Peripherals:] A collection of peripherals enables
      various use cases using the \SPIRS{} platform.

      In addition to DDR3 memory, the platform includes
      Ethernet, UART, SPI, JTAG, Boot ROM, PLIC, and CLINT\@.
\end{description}

\Paragraph{Prototype Platform Implementation}
A prototype of the \SPIRS{} platform integrates
the above components, and demonstrates
secure communication between the \RISCV{} processor,
the \RoT{}, and
the \gls{HLOS}, mediated by the \TEE{}.

To speed up and facilitate development, the \SPIRS{} \TEE{} \SDK{} also
includes a preconfigured virtualized environment, based on QEMU\@.

\subsection{\SPIRS{} TEE Platform}\label{subsec:spirs_tee_platform}

\begin{figure}[tb]
  \begin{center}
  \includegraphics[width=1.0\textwidth]{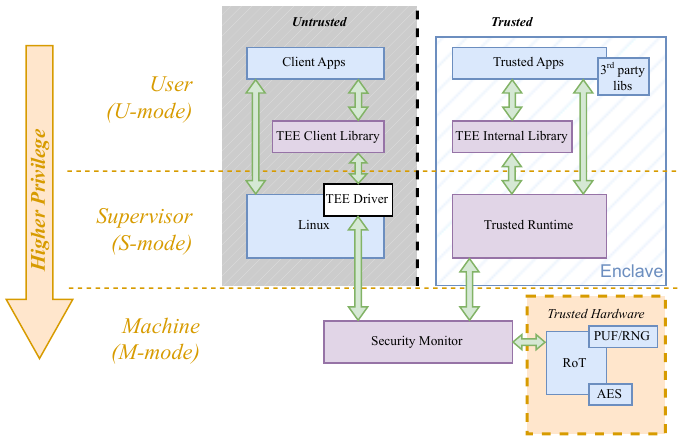}
  \end{center}
  \caption{Basic \SPIRS{} \TEE{} stack}%
  \label{fig:TEEArch}
\end{figure}

The \TEE{} is a crucial element of the
\SPIRS{} platform. It enables secure execution of critical operations,
isolating the trusted codebase from potentially compromised or
untrusted components.
\autoref{fig:TEEArch} visually summarizes the main components of the
\SPIRS{} \TEE{} architecture and their interactions.
The rest of this section, and the following one, provide more details
about these components and how different actors interact with the
\SPIRS{} platform.

At the core of the \TEE{} design is the \textbf{\SM{}},
which has the highest execution privilege on the platform.
The \SM{} manages memory and device access,
using \PMP{}~\cite[Section~3.7]{RV:Privileged} for compartmentalization
and leveraging the hardware \RoT{} to ensure security.
The \SM{} also mediates communication between \textbf{\TAs{}}
running in the \TEE{} and
untrusted components,
such as userspace \textbf{\CAs{}}
within the \gls{HLOS}.

\TAs{} are lightweight applications running within
the \TEE{}, with minimal permissions to reduce the attack surface.
These applications rely on the
\textbf{\SPIRS{} Trusted Runtime}
for accessing \RoT{} services and basic functionalities like memory
management.

\subsection{Software: SPIRS-TEE-SDK}\label{subsec:spirs_sw}

\begin{figure}[h!]
  \begin{center}
  \includegraphics[width=1.0\textwidth]{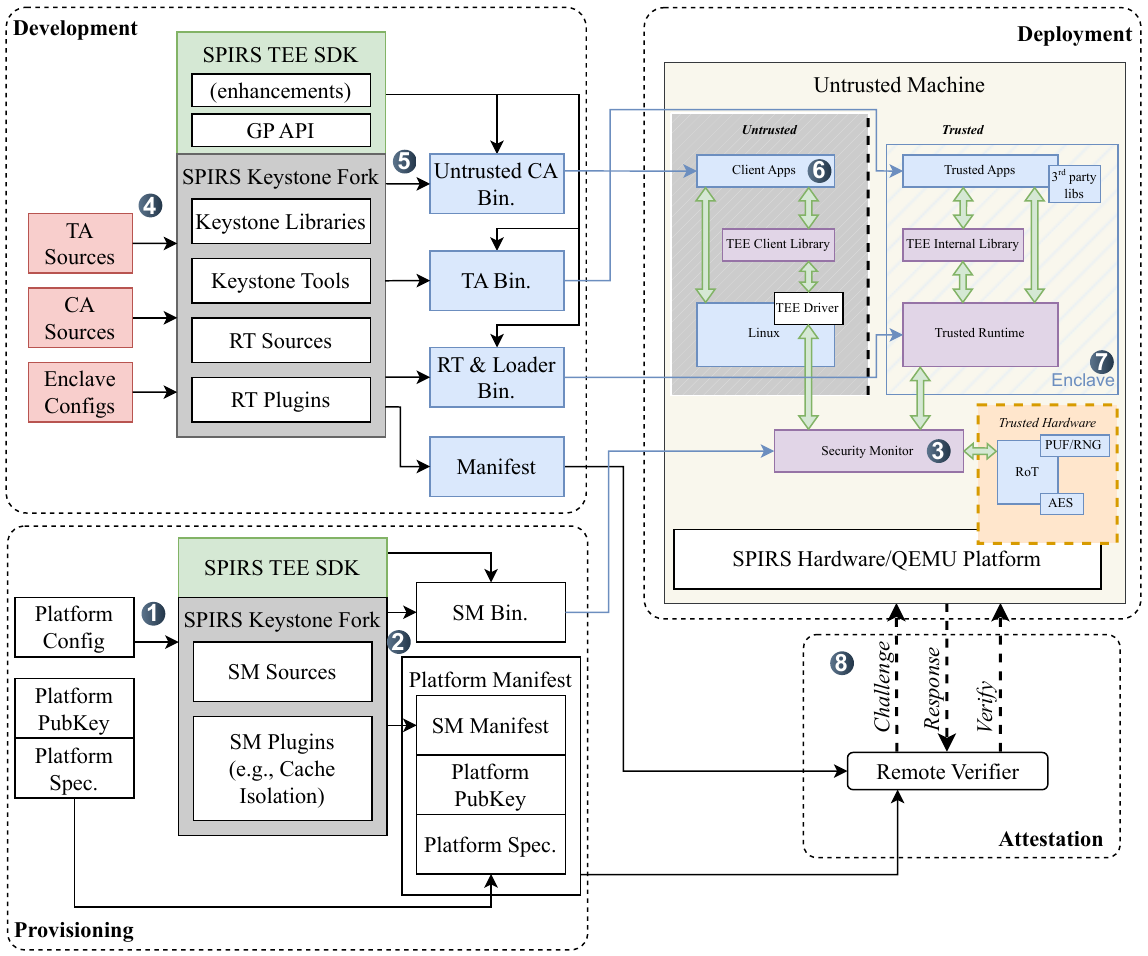}
  \end{center}
  \caption{\centering{\SPIRS{} TEE SDK Workflow}}%
  \label{fig_spirs_workflow}
\end{figure}

The \SPIRS{} \TEE{} \SDK{} simplifies the
development, testing, and deployment of applications leveraging the \SPIRS{} \TEE{}.
It is built upon a fork of the
open-source Keystone framework~\citep{DBLP:conf/eurosys/LeeKSAS20},
tailored specifically for \RISCV{} hardware
platforms.

The \SDK{} defines pairs of applications: \CAs{}
executing in the untrusted environment, and
\TAs{}
running securely within the \TEE{}.
Communication between \TAs{} and userspace \CAs{}
is mediated through dedicated
\textbf{\TEE{}~Client}~(\texttt{tee\_client\_api})
and internal (\texttt{tee\_internal\_api})
libraries,
based on the \GP{} \TEE{}
specifications~\citep{gp_specs,gp_client_api,gp_internal_api}.

Briefly, the typical workflow, depicted in \autoref{fig_spirs_workflow}
and detailed below,
involves writing application logic for both the \CA{} and \TA{},
configuring build parameters
(such as \texttt{APP\_NAME} and a unique \texttt{TA\_UUID}),
and compiling them alongside the
\SPIRS{} Trusted Runtime
into a single executable Keystone application package (\texttt{.ke}).

\definecolor{circleleft}{RGB}{112,133,155}
\definecolor{circleright}{RGB}{39,57,77}
\newcommand{\circnum}[1]{%
  \tikz[baseline=(num.base)]{ %
    \shade[shading=axis, left color=circleleft, right color=circleright]
      (0,0) circle (0.5em);
    \node[white, font=\sffamily\bfseries\footnotesize, inner sep=0pt]
      (num) at (0,0) {#1};
  }%
}

\autoref{fig_spirs_workflow} comprehensively depicts the different phases of a
typical workflow when using the
\SPIRS{} \TEE{} \SDK{}:

\begin{description}
  \item[Provisioning:] The vendor of the \SPIRS{} platform can provision
    specific boards, based on a description of the
    platform configuration~\circnum{1}.
    This is processed by the \SPIRS{} fork of Keystone, augmented
    with further tweaks provided by the \SPIRS{} \SDK{}, to
    generate~\circnum{2} a trusted \SM{} binary and a platform manifest,
    which embeds an \SM{} manifest authenticating the binary, together with a
    public key and a specification related to the platform.
    Furthermore, this step is also used to initialize the \SDK{} environment for
    development.
  \item[Development:] Once the \SDK{} environment is provisioned, developers can
    implement and maintain, for each of their \TAs{}, the code base~\circnum{4}
    for the \TA{}/\CA{} pair, alongside the configuration
    tuning the enclave which will be spawned to run a specific \TA{} in the
    \SPIRS{} \TEE{} (e.g., configuring memory boundaries and access to resources
    and peripherals, an \texttt{APP\_NAME} and a unique \texttt{TA\_UUID}).
    These will be processed through the \SPIRS{} fork of Keystone, enhanced by
    tweaks and APIs provided by the \SPIRS{} \SDK{}, to generate~\circnum{5}
    executable binary artifacts for the \CA{} and
    a single executable Keystone application package (\texttt{.ke}), which
    conveniently collects the \TA{} binary artifact, the \SPIRS{} Trusted
    Loader, the \SPIRS{} Trusted Runtime (based on the Keystone Trusted
    Runtime \textit{Eyrie}), and a manifest authenticating the bundle.

    The \SDK{} facilitates rapid prototyping and testing via integration with
    the QEMU virtualized environment, offering a practical alternative to
    physical hardware during early development stages.

    Additional capabilities include straightforward integration of third-party
    libraries via static linking, streamlined environment setup through provided
    scripts, and automated test execution within QEMU\@.
  \item[Deployment:] The operator of the platform can install the \SPIRS{} board
    where needed.
    They deploy on its storage,
    the \SM{} binary output~\circnum{2}
    obtained through the \emph{Provisioning},
    their choice of a
    supported Linux kernel and filesystem
    (the \SDK{} provisioning also produces these, although operators can
    optionally ship a different one, as needed),
    and on top of the filesystem they also install the
    \emph{Development} artifacts~\circnum{5}
    for the \CA{} and the executable Keystone application package.

    At boot, the board uses the provisioned public key to authenticate the
    \SM{} binary, before loading and executing it in memory~\circnum{3} to
    continue the Secure~Boot process: the Linux kernel and filesystem are hashed
    before loading, and their values saved in dedicated security registers.
    The Linux \gls{IMA} subsystem can read those values, and use additional
    security registers to record the digests of further loaded components.

    Once boot is completed, a kernel driver allows userspace
    to interact with the \SM{} to
    spawn an isolated \TEE{} enclave~\circnum{7} to run the \TA{} from the
    executable Keystone application package (\texttt{.ke}),
    as well as the associated \CA{}~\circnum{6} in the
    \emph{Untrusted}~\gls{HLOS} to interact with it.
  \item[Attestation:] Finally, a Remote Attestation protocol~\circnum{8}
    based on Challenge/Response, allows a Remote Verifier to attest the
    authenticity of the hardware and software components of the platform, based
    on the Platform and \TA{} manifests generated in \emph{Provisioning} and
    \emph{Development}.
    The attestation report can also leverage the Linux \gls{IMA}~subsystem and
    the secure registers, to let the \SM{} generate a signed quote covering all
    the measured components.

    The Remote Attestation mechanism integrated
    within the \SPIRS{} \TEE{} \SDK{}
    extends the security properties of the established communication channels.
    On top of the confidentiality provided by protocols such as \gls{TLS},
    upon successful completion of the Remote Attestation protocol,
    the communicating parties additionally gain
    \emph{mutual authentication}, guaranteeing that each party's identity has been
    securely verified;
    \emph{integrity}, ensuring not only the exchanged data but also the
    software state of the attested platform have not been tampered with,
    as well as proof of a genuine hardware configuration; and
    \emph{freshness}, confirming that the attestation data reflects the
    current state rather than a replayed past state.

    These extra security guarantees significantly reduce the threat surface
    protecting the communication channel against impersonation, replay, and
    tampering attacks, ultimately enhancing trust in the
    remotely attested execution environments.
\end{description}
 
\section{Use case: Entropy and IoT}

True randomness, or entropy---a reliable measure of randomness---can be achieved with methods that can generate actual random data.
This entropy is hard to achieve as it is not straightforward to identify methods for gathering truly random inputs or generating truly random data using specific input sources.
The introduction describes a variety of previous methods for achieving entropy.
However, these methods can call true randomness into question if the methods or inputs follow a pattern or the data can be manipulated.

Methods used for generating true randomness are known as \TRNGs{}.
\TRNGs{} include, but are not limited to, chaos~\citep{DBLP:books/sp/14/StipcevicK14}, electrical noise~\citep{DBLP:books/sp/14/StipcevicK14},
free-running oscillators~\citep{DBLP:books/sp/14/StipcevicK14}, quantum effects~\citep{DBLP:books/sp/14/StipcevicK14},
and radioactive decay~\citep{you2023information}.
Chaos is addressed using methods like biosignals~\citep{DBLP:journals/iot/ClementeLopezRM24}, multi-dimensional chaotic systems~\citep{DBLP:journals/iot/FanW24}, and wind flow~\citep{Kim2024}.
To explain further, radioactive decay is used to retrieve random signals to support creating true randomness~\citep{you2023information}.
\TRNGs{} based on Ring Oscillators are a recent and well-known technique for achieving entropy in \IoT{} systems~\citep{DBLP:conf/dcis/Rojas-MunozSMB22}.

In addition to \TRNGs{} that use true randomness,
there are also \PRNGs{} which are algorithms that transform input seed values
into longer and uniformly distributed deterministic bit streams,
meaning that the same seed value will always result in the same output value.
In order for a \PRNG{} to be considered a \CSPRNG{},
it must provide both forward and backward unpredictability.

Because of their limited resources, constrained \IoT{} devices may encounter difficulty in producing the strong entropy
necessary for cryptographically secure random numbers~\citep{DBLP:journals/csur/KietzmannSW21}.
Furthermore, Petro and Cecil~\citep{defconiotrng} identify numerous additional security issues with \IoT{} \RNGs{}, including
(i) Lack of access to proper \CSPRNG{} subsystems due to lack of modern OS\@;
(ii) Developers frequently making incorrect calls to hardware \TRNGs{};
(iii) The difficulty of handling hardware \TRNG{} errors correctly;
(iv) Inadequate documentation for bare-metal function calls; and
(v) Flaws in the entropy distillation process that generate subpar quality hardware \TRNG{}.

A way to solve the problem of limited entropy for \IoT{} devices is to use \EaaS{}~\citep{NistEaaSProject},
first introduced by Vassilev and Staples in 2016~\citep{DBLP:journals/computer/VassilevS16}.
In contrast to \emph{Randomness Beacons}~\citep{NistRandomnessBeaconProject},
in an \EaaS{} system, a server only distributes secure and fresh entropy to clients that specifically request it,
instead of broadcasting it in timed intervals.
Previous research demonstrated that \TEE{} offers security for diverse applications,
and we are now broadening this research domain by introducing an external entropy provision service~\citep{DBLP:conf/IEEEares/PajuJNSMB23}.
The entropy received from an \EaaS{} server with secure \TEE{} hardware is inherently suitable for cryptographic use,
unlike the publicly known entropy provided by the more-researched Randomness Beacons.
\section{\EaaS{} Design}\label{sec:design}

Our \EaaS{} solution consists of three main components:
(i) an \IoT{} client;
(ii) a \TES{}; and
(iii) entropy sources.
The high-level idea is that a resource-constrained \IoT{} device sends a \HTTP{} request to a server,
which subsequently processes the request by fetching fresh entropy from the entropy sources,
generates strong entropy by combining the entropy received from the entropy sources,
and finally transmits the entropy to the \IoT{} client.
The protocol incorporates a verification function that validates both
the digital signature generated by the \TES{}
and the freshness of the received entropy.

We employ a fleet of \IoT{} devices in our implementation,
each of which has a limited capacity to generate entropy.
However, none of them has sufficient resources to independently generate strong and crypto-secure entropy.
Thus, the \IoT{} client is represented by one of the IoT devices in the fleet,
while the remaining devices serve as the entropy sources.

Because the entropy fetching and generation portion of the protocol is executed within a \TEE{},
the trust is placed in the \TEE{} technology rather than a third-party entity (server operator).
Thus, we refer to this server as a \TES{} instead of a \TTP{} server.

In order to use asymmetric encryption,
we assume a key provisioning model which gets rid of the chicken-and-egg problem;
i.e., we assume that the manufacturer has provided a strong initial key to the \IoT{} client.
Thus, the \IoT{} client does not need to use \EaaS{} for
generating the strong key required for using the \EaaS{} system in the first place.
Additionally, we assume that the \IoT{} client has previously obtained the public key of the \TES{} ($pk_{TES}$).

When the \IoT{} client wants to request entropy from the \TES{},
it first generates a timestamp ($t_{1}$) and then sends a self-signed request
to the \TES{} that has been encrypted with the public key of the \TES{} ($pk_{TES}$).
The encrypted request contains the public key ($pk_{IoT}$), the quantity of entropy requested ($\Delta S$),
and a digital signature ($\sigma_1$). %

After the \TES{} receives the request,
it fetches an amount $\Delta S$ of entropy from the entropy sources (denoted by $\Delta S_{1}$ and $\Delta S_{2}$),
and then combines the entropy (denoted by $S$) and generates a timestamp ($t_{2}$).
\TES{} also creates a fixed-length symmetric session key ($sk_{session}$).
To distribute the symmetric key to the \IoT{} client,
the \TES{} encrypts the symmetric key $sk_{session}$ with the public key of the \IoT{} client $pk_{IoT}$.
This symmetric key $sk_{session}$ is used to encrypt the encapsulated message that contains both
the generated entropy $S$ and the timestamp $t_{2}$.
Finally, \TES{} signs both the encrypted symmetric session key and the encapsulated message
with its secret key ($sk_{TES}$). The result is a digital signature ($\sigma_2$). %

The response from \TES{} to the \IoT{} client is thus a message that contains:
(i) a fixed-length symmetric session key $sk_{session}$, encrypted with the public key of the \IoT{} client $pk_{IoT}$;
(ii) an encapsulated response to the request for entropy.
This response contains the generated entropy $S$ and the timestamp $t_{2}$
and is encrypted with the fixed-length symmetric session key $sk_{session}$;
(iii) a digital signature $\sigma_2$, signed with the secret key of \TES{} $sk_{TES}$. %

\begin{figure}[tb]
  \begin{center}
  \includegraphics[width=1.0\textwidth]{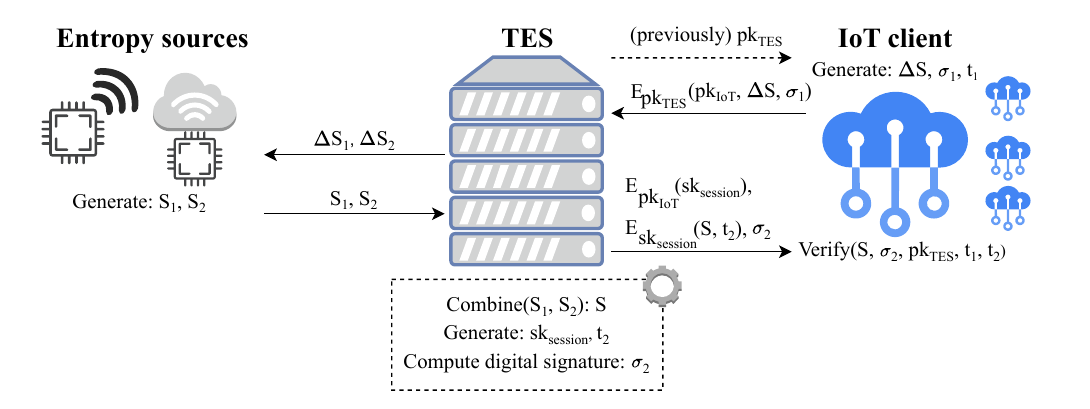}
  \end{center}
  \caption{\centering{\EaaS{} Protocol}}%
  \label{fig_eaas_protocol}
\end{figure}

After decrypting the message, the \IoT{} client verifies that:
(i) the message contains the requested entropy $S$ with the requested amount of entropy $\Delta S$,
the digital signature $\sigma_2$, and the timestamp of the entropy generation $t_{2}$; %
(ii) the freshness of the generated entropy (${t_2} > t{_1}$); and
(iii) the digital signature $\sigma_2$ is valid. %
The protocol is depicted in \autoref{fig_eaas_protocol}.

\subsection{Security considerations and evaluations}\label{subsec:considerations}
\Paragraph{\TEE{} provides security even if the server operator is malicious}
The \TEE{} technology ensures the security of the \EaaS{} system,
even in the event of a malicious server operator,
provided it functions properly.
This justifies calling the server a \TES{} rather than a \TTP{} server.
Side-channel attacks are a potential attack vector against the \TEE{} technology
that may undermine the security assumptions.
While we are currently unaware of any side-channel attacks against the
\RISCV{}-based \TEE{} we use in our \EaaS{} implementation,
it is important to be aware of this potential attack vector.

\Paragraph{Self-signing the request for entropy protects against a malicious actor on the communication link}
We assume that there is a malicious actor on the link or that the link may have some corruption.
Self-signing the request with the \IoT{} device's private key binds the public and private keys,
as well as the entropy quantity request for the transaction.
Similar self-signing is used in \PKI{} with a Certificate Signing Request
to certify a public key by a Certificate Authority.
Without this self-signing, a malicious actor could perform a simple \DoS{} attack
by modifying the public key, preventing the requester from decrypting the response.
Additionally, the malicious actor could alter the quantity of requested entropy,
resulting in insufficient entropy for the requesting \IoT{} client.
Of course, a malicious actor can simply drop messages, but this only affects the requester.
If the message is returned and rejected by the requester,
it is likely that they will request it again.
This could impact the \TES{}, either by keeping it busy servicing constant messages or,
in the worst-case scenario, by depleting the entropy source, which can affect all IoT{} nodes.

\Paragraph{Throttling requests from a single source stops the depletion of entropy}
To avoid entropy depletion, we use a throttling mechanism on the \TES{}
to limit the number of requests from a single source.
This protects the \EaaS{} system from a bad actor who repeatedly
makes valid requests to the \TES{} with the goal of consuming all entropy.
In addition to malicious actors, this will also offer protection against
misconfigured or buggy networks/devices that repeatedly request entropy for non-malicious reasons.
An alternative solution to fixing this problem would be to assume
that the initial factory provisioning provides not only a key-pair
but also a certificate that can be verified by the \TES{}.
However, this approach adds more overhead to the protocol and
may not even be feasible on these constrained \IoT{} devices.
Thus, we instead implement a throttling mechanism.

\Paragraph{A symmetric session key is used for encrypting the \TES{} response
because of public key encryption data size limitations}
The payload size for public key encryption is limited.
For example, RSA can only perform one encryption step,
which means it can only encrypt data up to the key size minus padding and header data.
After removing padding, header data, timestamp, and signature data from our \TES{} response,
we are left with the data size available for the actual entropy payload.
This means that either the requested amount of entropy must be less than this amount,
or the protocol must support multiple messages in order to return all of the requested entropy
in case the requested amount exceeds the available data size.
Thus, to avoid unnecessary protocol overhead,
we use fixed-length symmetric session keys to encrypt the \TES{} response payload.
\section{Software implementation}\label{sec:results}
We implemented our \EaaS{} solution using the \SPIRS{} \TEE{} \SDK{} with the QEMU virtualized environment.
The software implementation was developed and tested on an Ubuntu 22.04.4 LTS machine.
We used the QEMU platform instead of actual SPIRS hardware due to the ease of implementation and speed of development.
The QEMU platform contains a \RISCV{} virtual machine image that includes bootloaders, \SM{}, Linux kernel, and rootfs.

To provide developers with easy access to cryptographic functions, we used OpenSSL as a third-party library.
In order to get the third-party library to work, we built a static OpenSSL library with a RISC-V musl toolchain
and added it to the \SPIRS{} \TEE{} \SDK{} build process.

We used a minimal \TCB{} design, running only the necessary components inside the TEE enclave (\TA{}).
In practice, only the code for obtaining and combining entropy,
generating the keys, computing and verifying the digital signatures, and
encrypting and decrypting the messages is run inside a \TA{}.
The rest of the code is executed on the \CA{}.

\NTP{} is used for timestamps.
AES-128 is used for symmetric encryption and
RSA-3072 is used for asymmetric encryption.
We use a hybrid scheme: RSA-3072 for signatures and AES-128 for payload encryption.

The \EaaS{} demo application serves as a functional Proof-of-Concept illustrating
that the \SPIRS{} \TEE{} \SDK{} is suitable for the practical development of useful \TAs{}.

\subsection{Limitations}\label{subsec:limitations}
A limitation of our work is that we are only using entropy gathered
from the entropy sources, which in our case are the \IoT{} devices in the fleet,
all of which are based on the \PUF{}/\TRNG{} inside the hardware \RoT{} of \SPIRS{} devices.
Therefore, any potential flaws with this \PUF{}/\TRNG{} implementation would also directly affect
the quality of entropy generated by our \EaaS{} system.
To address this problem, our \EaaS{} implementation could be modified to also
use other trusted entropy sources.
We recommend that real-world \EaaS{} implementations use many trusted entropy sources
that are based on different phenomena,
so that any potential flaws in one method do not invalidate the quality of generated entropy.
Since our \EaaS{} design allows the easy addition of entropy sources,
this is easily achievable in any real-world implementation based on our design.

The development happened on a virtualized QEMU environment
instead of the actual SPIRS hardware.
To get the software working on the actual SPIRS hardware,
we would need to develop further mappings between the OpenSSL library
and the trusted hardware implementations.
However, since the cryptographic functions used in the software implementation
are mostly the same functions that are supported by the trusted hardware components of the
\SPIRS{} \TEE{} hardware platform, this should be straightforward to achieve.
Future research should include hardware and performance benchmarking
to assess the practicality and usability of our solution.

Our implementation relies on a single \TES{}, creating a potential single point of failure.
However, our design could be easily extended to include features such as load balancing
for high availability in production environments where a single point of failure is not acceptable.
The design could also be extended for a decentralized solution.

The threats posed by a future \gls{CRQC} inherently affect this project
as well.
In this work, we did not directly address \gls{PQC} concerns, as the
efforts for the \gls{PQC} transition are orthogonal to our application.
Nonetheless, it is worth remarking that as both confidentiality and
authentication based on asymmetric cryptography are impacted by the
quantum threats, real-world deployments of this work should select
\gls{PQ/T}~hybrid algorithms to secure the communication
channels against attacks such as \gls{HNDL}.
While threats to authentication do require an active attacker with
access to a \gls{CRQC}, the direct threat here does not need to be
addressed as urgently as asymmetric encryption.
Nonetheless, as we are well aware of the many challenges
in any migration affecting the \PKI{}, we remark on the
urgency to address the \gls{PQC} transition of the \PKI{} to provide
a sustainable migration path for devices and applications such as the
ones described here.
Finally, on a related note, we recognize that one limitation of the current
implementation of the \SPIRS{} platform is that its \RoT{} does not
embed support for \gls{PQC} primitives: a natural evolution for the
\SPIRS{} project would consist in providing support in hardware and
firmware to establish a chain of trust based on \gls{PQ/T}~hybrid
algorithms.
\section{Conclusion}\label{sec:conclusion}

In this paper, we presented a working \EaaS{} demo application and its design
that utilizes the \SPIRS{} \TEE{} \SDK{},
demonstrating that developing useful \TAs{} for an open-source \RISCV{}-based
\TEE{} is both feasible and effective.
Our implementation highlights the potential of combining open hardware with secure software to address real-world challenges in \IoT{} security.

\Paragraph{\EaaS{} warrants further research}
Our work showcases that designing and building a working \EaaS{}
implementation is practical
and useful for \IoT{} devices that lack sufficient resources
to generate strong entropy by themselves.
However, there is limited scientific research
and practical applications for \EaaS{},
as most of the attention seems to be on the somewhat-related Randomness Beacons.
We believe our implementation illustrates that \EaaS{}, particularly when backed by a \TEE{} for a \TES{},
offers distinct advantages in selective entropy delivery, remote attestation, and system scalability.
\EaaS{} ensures confidentiality, freshness, and trust in entropy distribution---making it a compelling direction for both academic inquiry and industrial adoption.
We encourage future work to explore alternative entropy sources, multi-tenant \TES{} infrastructures, and deployment in heterogeneous hardware ecosystems.

\Paragraph{The \SPIRS{} \TEE{} \SDK{} allows practical development of \TAs{} for open-source \RISCV{} hardware}
Our \EaaS{} demo application demonstrates that the
\SPIRS{} \TEE{} \SDK{} enables practical development of \TAs{}
for the \RISCV{}-based \SPIRS{} hardware platform.
This application is a meaningful demonstration of how \TAs{} can be developed
on open-source \RISCV{}-based \TEE{} hardware.
The open-source toolchain, support for virtualization via QEMU,
and integration of cryptographic primitives through trusted runtime components
make the \SDK{} suitable for rapid prototyping and academic experimentation.
We anticipate that the \SPIRS{} platform and SDK will lower the barrier
to entry for developing new \TAs{} in diverse domains, including embedded security, supply chain attestation, and privacy-preserving analytics.

\Paragraph{Our design warrants benchmarking and comparison with other solutions}
This paper presents our \EaaS{} design and its simple QEMU-based implementation.
The scope did not include performance or hardware measurements.
Future research should measure the overhead of the \EaaS{} implementation and evaluate its scalability for multiple concurrent \IoT{} clients.
Additionally, we recommend comparing our solution to other \EaaS{} and Randomness Beacon systems.
\section{Code availability}\label{sec:code}

As we fully support Open Science principles,
we have published our code with an open-source (MIT) license.
The code can be accessed through the following Git repository:
\url{https://gitlab.com/nisec/spirs-tee-sdk-eaastee-public-fork}.
The Git repository includes a detailed \emph{README.md} file with a
tutorial for using the \SPIRS{} \TEE{} \SDK{}
as well as running our demo \EaaS{} application.
The \emph{README.md} file also explains the chosen algorithms
and design choices in further detail.
 
\ifANON\else

\section{Authorship Contributions}\label{sec:contributions}

J.N., A.A., N.T., and B.M. acquired funding.\\
A.P. led the article writing and administration.\\
A.C.A. and N.T. designed the RISC-V SPIRS SDK.\\
A.P. and J.N. designed the entropy provision service.\\
J.S. analyzed the entropy provision protocol.\\
A.A., N.T., and B.M. validated and verified our work.\\
M.K. contributed to the visualization of diagrams.\\
A.P., J.N., and B.M. proofread the manuscript.\\
B.M., A.P., and J.N. edited the final version of the paper.\\
Contributor Roles Taxonomy (\href{https://credit.niso.org/}{CRediT}) statement of authorship contribution.

\begin{table}[htb!]
\iftrue%
  \resizebox{1.0\linewidth}{!}{%
    \begin{tabular}{
      p{0.34\linewidth}
      >{\centering\arraybackslash}p{0.08\linewidth}
      >{\centering\arraybackslash}p{0.08\linewidth}
      >{\centering\arraybackslash}p{0.10\linewidth}
      >{\centering\arraybackslash}p{0.08\linewidth}
      >{\centering\arraybackslash}p{0.08\linewidth}
      >{\centering\arraybackslash}p{0.08\linewidth}
      >{\centering\arraybackslash}p{0.08\linewidth}
    }
      \toprule
      \textbf{Contributions} & \textbf{A.P.} & \textbf{J.N.} & \textbf{A.C.A.} & \textbf{N.T.} & \textbf{J.S.} & \textbf{M.K.} & \textbf{B.M.} \\
      \midrule
              Data curation  & $\bullet$     & $\circ$       & $\circ$         & $\circ$       & $\circ$       & $\circ$       & $\circ$ \\
             Formal Analysis & $\bullet$     & $\circ$       & $\circ$         & $\bullet$     & $\circ$       & $\circ$       & $\circ$ \\
         Funding acquisition & $\circ$       & $\bullet$     & $\bullet$       & $\bullet$     & $\circ$       & $\circ$       & $\bullet$ \\
               Investigation & $\bullet$     & $\bullet$     & $\circ$         & $\bullet$     & $\bullet$     & $\circ$       & $\circ$ \\
                 Methodology & $\bullet$     & $\bullet$     & $\circ$         & $\bullet$     & $\bullet$     & $\circ$       & $\bullet$ \\
      Project administration & $\bullet$     & $\bullet$     & $\bullet$       & $\circ$       & $\circ$       & $\circ$       & $\circ$ \\
                   Resources & $\circ$       & $\circ$       & $\bullet$       & $\circ$       & $\circ$       & $\circ$       & $\circ$ \\
                    Software & $\bullet$     & $\bullet$     & $\bullet$       & $\bullet$     & $\circ$       & $\circ$       & $\circ$ \\
                 Supervision & $\bullet$     & $\bullet$     & $\bullet$       & $\circ$       & $\circ$       & $\circ$       & $\circ$ \\
                  Validation & $\circ$       & $\circ$       & $\circ$         & $\bullet$     & $\circ$       & $\circ$       & $\circ$ \\
               Visualization & $\bullet$     & $\bullet$     & $\bullet$       & $\bullet$     & $\circ$       & $\bullet$     & $\circ$ \\
    Writing---original draft & $\bullet$     & $\bullet$     & $\bullet$       & $\bullet$     & $\bullet$     & $\circ$       & $\circ$ \\
 Writing---review \& editing & $\bullet$     & $\bullet$     & $\circ$         & $\circ$       & $\circ$       & $\circ$       & $\bullet$ \\
      \bottomrule
      \end{tabular}%
}
\else
  \Huge\textbf{PLACE\\HOLDER}
\fi
\end{table}

\FloatBarrier{}
 \fi

\ifANON\else
\section{Acknowledgments}\label{sec:acknowledgments}

This work was supported by the European Commission under the Horizon Europe funding programme, %
as part of the projects SafeHorizon (Grant Agreement 101168562) and QUBIP (Grant Agreement 101119746). %
Views and opinions expressed are, however, those of the author(s) only and do not necessarily reflect those of the European Union. %
Neither the European Union nor the granting authority can be held responsible for them.
The authors declare no competing interests.
 \fi

\ifIACRTRANS{}
\printbibliography{}

\begin{thebibliography}{10}
\providecommand{\url}[1]{\texttt{#1}}
\providecommand{\urlprefix}{URL }
\providecommand{\doi}[1]{https://doi.org/#1}

\bibitem{DBLP:journals/corr/abs-2208-11935}
Amil, A., Gupta, S.: A universal whitening algorithm for commercial random
  number generators. CoRR  \textbf{abs/2208.11935} (2022),
  \url{https://doi.org/10.48550/arXiv.2208.11935}

\bibitem{RV:Privileged}
Asanovic, K., Ashenden, P., Avižienis, R., Bachmeyer, J., Baum, A.J., Behrens,
  J., Bonzini, P., Bukin, R., Celio, C., Chang, C., Chisnall, D., Coulter, A.,
  Dabbelt, P., Dalrymple, M., Donahue, P., Favor, G., Ferguson, D., Gauthier,
  M., Glew, A., Guo, G., Frysinger, M., Hauser, J., Horner, D., Johansson, O.,
  Kruckemyer, D., Lee, Y., Lustig, D., Lutomirski, A., Mundkur, P.,
  Neuschäfer, J., Nikhil, R., O’Rear, S., Ou, A., Ousterhout, J., Patterson,
  D., Pavlov, D., Phillips, K., Scheid, J., Schmidt, C., Taylor, M., Terpstra,
  W., Thomas, M., Thorn, T., VanDeWalker, R., Wachs, M., Wallach, S., Waterman,
  A., Wolf, C., Zandijk, R.: The {RISC-V} Instruction Set Manual. {Volume II}:
  Privileged Architecture. RISC-V International (2024),
  \url{https://github.com/riscv/riscv-isa-manual/releases/tag/20240411},
  version 20240411.

\bibitem{DBLP:journals/iot/ClementeLopezRM24}
Clemente{-}L{\'{o}}pez, D., de~Jesus~Rangel{-}Magdaleno, J.,
  Mu{\~{n}}oz{-}Pacheco, J.M.: A lightweight chaos-based encryption scheme for
  {IoT} healthcare systems. Internet of Things  \textbf{25},  101032 (2024),
  \url{https://doi.org/10.1016/j.iot.2023.101032}

\bibitem{CCPP84}
{EUROSMART}: {BSI-CC-PP-0084-2014: Security IC Platform Protection Profile with
  Augmentation Packages}.
  \url{https://www.bsi.bund.de/SharedDocs/Zertifikate_CC/PP/aktuell/PP_0084.html}
  (2014), {Bundesamt für Sicherheit in der Informationstechnik (BSI).
  Accessed: 2025-06-04}

\bibitem{DBLP:journals/iot/FanW24}
Fan, S., Wang, J.: Multi-dimension-precision chaotic encryption mechanism for
  {Internet of Things}. Internet of Things  \textbf{26},  101202 (2024),
  \url{https://doi.org/10.1016/j.iot.2024.101202}

\bibitem{gp_client_api}
{GlobalPlatform Inc.}: {TEE Client API Specification v1.0}. Tech. Rep.
  {GPD\_SPE\_007}, {GlobalPlatform Inc.} (2010),
  \url{https://globalplatform.org/specs-library/tee-client-api-specification/}

\bibitem{gp_internal_api}
{GlobalPlatform Inc.}: {TEE Internal Core API Specification v1.1.2}. Tech. Rep.
  {GPD\_SPE\_010}, {GlobalPlatform Inc.} (2016),
  \url{https://globalplatform.org/specs-library/tee-internal-core-api-specification/}

\bibitem{gp_specs}
{GlobalPlatform Inc.}: {GlobalPlatform TEE Specification Library}.
  \url{https://globalplatform.org/specs-library/?filter-committee=tee} (2025),
  accessed: 2025-06-04

\bibitem{DBLP:journals/csur/KietzmannSW21}
Kietzmann, P., Schmidt, T.C., W{\"{a}}hlisch, M.: {A Guideline on Pseudorandom
  Number Generation (PRNG) in the IoT}. {ACM} Comput. Surv.  \textbf{54}(6),
  112:1--112:38 (2021), \url{https://doi.org/10.1145/3453159}

\bibitem{Kim2024}
Kim, M.S., Tcho, I.W., Choi, Y.K.: Cryptographic triboelectric random number
  generator with gentle breezes of an entropy source. Scientific Reports
  \textbf{14}(1), ~1358 (Jan 2024),
  \url{https://doi.org/10.1038/s41598-024-51939-2}

\bibitem{DBLP:conf/crypto/Krawczyk10}
Krawczyk, H.: {Cryptographic Extraction and Key Derivation: The HKDF Scheme}.
  In: Rabin, T. (ed.) Advances in Cryptology - {CRYPTO} 2010, 30th Annual
  Cryptology Conference, Santa Barbara, CA, USA, August 15-19, 2010.
  Proceedings. Lecture Notes in Computer Science, vol.~6223, pp. 631--648.
  Springer (2010), \url{https://doi.org/10.1007/978-3-642-14623-7\_34}

\bibitem{DBLP:conf/eurosys/LeeKSAS20}
Lee, D., Kohlbrenner, D., Shinde, S., Asanovic, K., Song, D.: {Keystone}: an
  open framework for architecting trusted execution environments. In: EuroSys.
  {ACM} (2020). \doi{10.1145/3342195.3387532}

\bibitem{NistRandomnessBeaconProject}
{National Institute of Standards and Technology (NIST)}: {Interoperable
  Randomness Beacons}.
  \url{https://csrc.nist.gov/projects/interoperable-randomness-beacons} (2011),
  accessed: 2025-06-04

\bibitem{NistEaaSProject}
{National Institute of Standards and Technology (NIST)}: {Entropy as a
  Service}. \url{https://csrc.nist.gov/Projects/Entropy-as-a-Service} (2016),
  accessed: 2025-06-04

\bibitem{VonNeumann}
von Neumann, J.: John von Neumann Collected Works, vol. 5: Design of Computers,
  Theory of Automata and Numerical Analysis, chap. Various Techniques Used in
  Connection with Random Digits, pp. 768--770. Pergamon Press, Oxford, England
  (1961), notes by George E. Forsythe and reproduced from J. Res. Nat. Bus.
  Stand. Appl. Math. Series, vol. 3, pp. 36--38 (1955).

\bibitem{DBLP:conf/IEEEares/PajuJNSMB23}
Paju, A., Javed, M.O., Nurmi, J., Savim{\"{a}}ki, J., McGillion, B., Brumley,
  B.B.: {SoK: A Systematic Review of TEE Usage for Developing Trusted
  Applications}. In: Proceedings of the 18th International Conference on
  Availability, Reliability and Security, {ARES} 2023, Benevento, Italy, 29
  August 2023- 1 September 2023. pp. 34:1--34:15. {ACM} (2023),
  \url{https://doi.org/10.1145/3600160.3600169}

\bibitem{defconiotrng}
Petro, D., Cecil, A.: {You're Doing IoT RNG} (2021),
  \url{https://bishopfox.com/blog/youre-doing-iot-rng}, {DEF CON 29}

\bibitem{DBLP:conf/dcis/Rojas-MunozSMB22}
Rojas{-}Mu{\~{n}}oz, L.F., S{\'{a}}nchez{-}Solano, S.,
  Mart{\'{\i}}nez{-}Rodr{\'{\i}}guez, M.C., Brox, P.: {True Random Number
  Generator based on RO-PUF}. In: 37th Conference on Design of Circuits and
  Integrated Systems, {DCIS} 2022, Pamplona, Spain, November 16-18, 2022.
  pp.~1--6. {IEEE} (2022), \url{https://doi.org/10.1109/DCIS55711.2022.9970032}

\bibitem{DBLP:conf/focs/SanthaV84}
Santha, M., Vazirani, U.V.: {Generating Quasi-Random Sequences from
  Slightly-Random Sources (Extended Abstract)}. In: 25th Annual Symposium on
  Foundations of Computer Science, West Palm Beach, Florida, USA, 24-26 October
  1984. pp. 434--440. {IEEE} Computer Society (1984),
  \url{https://doi.org/10.1109/SFCS.1984.715945}

\bibitem{spirs_web}
{SPIRS}: {SPIRS homepage}. \url{https://www.spirs-project.eu} (2025), accessed:
  2025-06-04

\bibitem{DBLP:books/sp/14/StipcevicK14}
Stipcevic, M., Ko{\c{c}}, {\c{C}}.K.: {True Random Number Generators}. In:
  Ko{\c{c}}, {\c{C}}.K. (ed.) Open Problems in Mathematics and Computational
  Science, pp. 275--315. Springer (2014),
  \url{https://doi.org/10.1007/978-3-319-10683-0\_12}

\bibitem{DBLP:journals/computer/VassilevS16}
Vassilev, A., Staples, R.: {Entropy as a Service: Unlocking Cryptography's Full
  Potential}. Computer  \textbf{49}(9),  98--102 (2016),
  \url{https://doi.org/10.1109/MC.2016.275}

\bibitem{you2023information}
You, I., Youn, T.: Information Security Applications: 23rd International
  Conference, WISA 2022, Jeju Island, South Korea, August 24--26, 2022, Revised
  Selected Papers. Lecture Notes in Computer Science, Springer Nature
  Switzerland (2023), \url{https://books.google.fi/books?id=K-irEAAAQBAJ}

\end{thebibliography}
\else

\fi

\end{document}